\documentstyle[epsfig]{elsart}
\begin{document}
\begin{frontmatter}

\title{The nonleptonic decays \boldmath{$B^+_c\to D^+_s \overline{D^0}$} and  
\boldmath{$B^+_c\to D^+_s D^0$} in a relativistic quark model.}

\author[jinr]{M.A. Ivanov,}
\author[mainz]{J.G. K\"{o}rner,}
\author[mainz,samara]{O.N. Pakhomova}
\address[jinr]{Bogoliubov Laboratory of Theoretical Physics, \\
Joint Institute for Nuclear Research, 141980 Dubna, Russia}
\address[mainz]{Institut f\"{u}r Physik, Johannes Gutenberg-Universit\"{a}t, 
D-55099 Mainz, Germany}
\address[samara]{Samara State University, 443011 Samara, Russia}

\begin{abstract}
In the wake of exploring $CP$-violation in the decays
of $B$ and $B_c$ mesons, we perform the straightforward 
calculation of their nonleptonic decay rates
within a relativistic quark model. We confirm that
the decays $B_c\to D_s \overline{D^0}$ and $B_c\to D_s D^0$
are well suited to extract the Cabibbo-Kobayashi-Maskawa
angle $\gamma$ through the amplitude relations because their
decay widths are the same order of magnitude. In the $b-c$ sector
the decays $B\to D K$ and $B_c\to D D$ lead to squashed triangles
which are therefore not so useful to determine the angle $\gamma$
experimentally. We also determine the rates for other nonleptonic
$B_c$-decays and compare our results with the results of other studies. 

\end{abstract}
       
\end{frontmatter}

As was pointed out in \cite{Masetti:1992in} and 
\cite{Fleischer:2000pp,Fleischer:2002nt} 
the decays $B^+_c\to D^+_s D^0(\overline{ D^0})$
are well suited for an extraction of the CKM angle $\gamma$ through 
amplitude relations. These decays are better suited for the extraction
of $\gamma$ than the similar decays of the $B_u$ and $B_d$ mesons 
because  the triangles in latter decays are very squashed. 
The $B_c$ meson has been observed
by the CDF Collaboration \cite{Abe:1998bc} in the decay
$B_c\to J/\psi l \nu$.  One could expect around $5\times 10^{10}$
$B_c$ events per year at LHC \cite{Gouz:2002kk} which gives us hope 
to use the $B_c$ decay modes for the  studying  $CP$-violation. 

\begin{figure}[t]
\begin{center}
\begin{tabular}{c}
\epsfig{figure=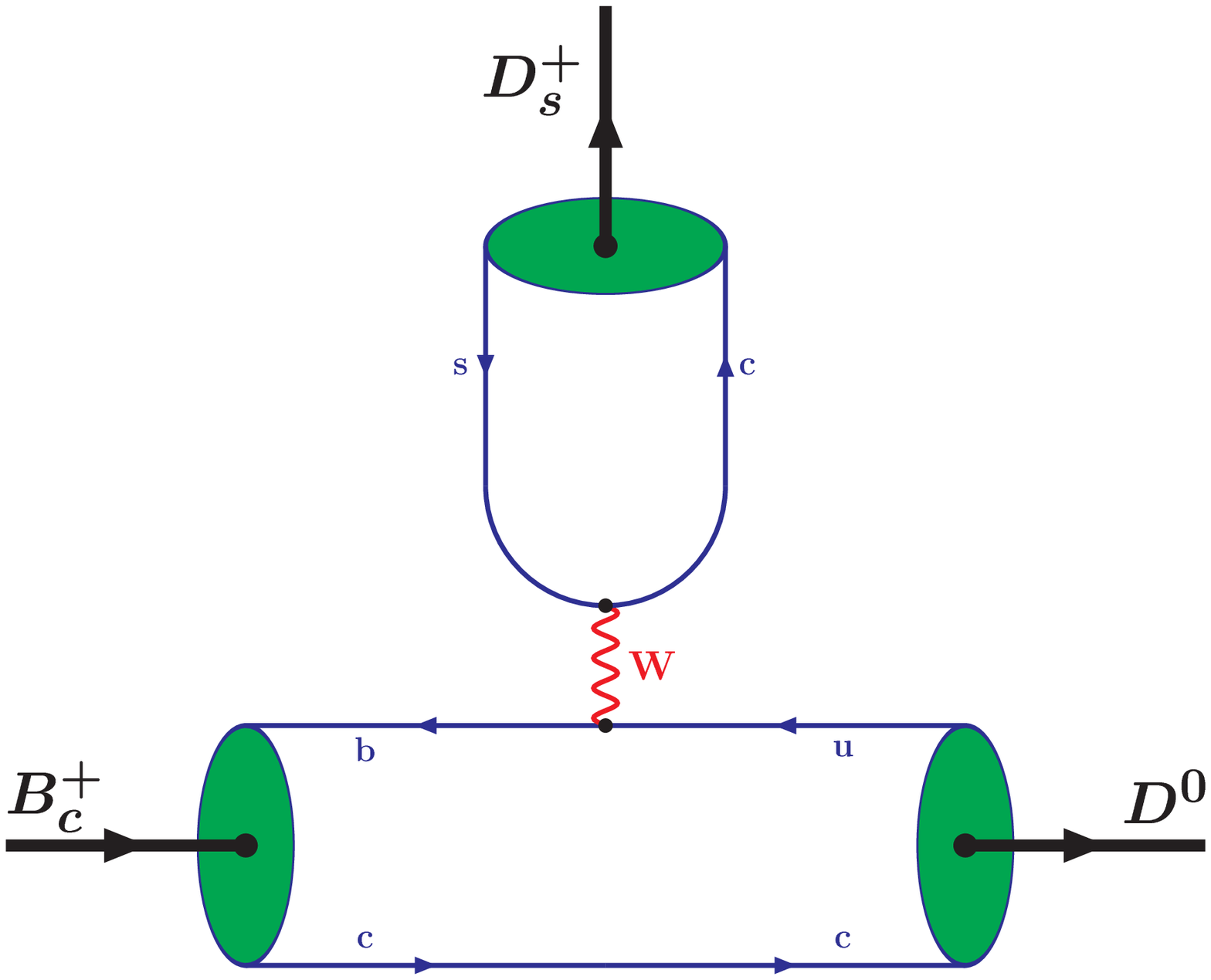,height=10cm} \\
\epsfig{figure=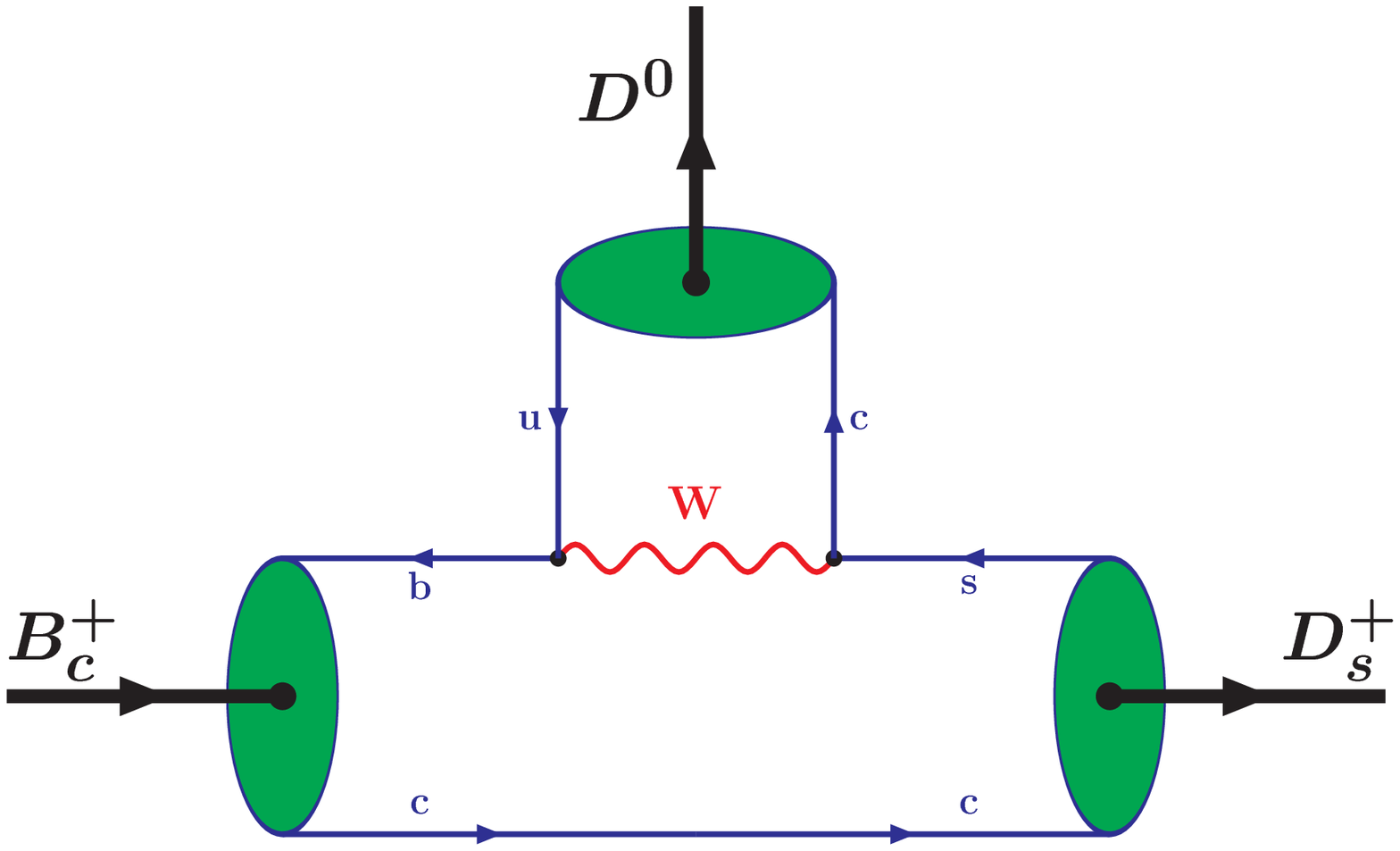,height=9cm}
\end{tabular}
\end{center}
\caption{Diagrams describing the decay $B_c\to D_s D^0$.}
\label{fig:bc_dsd}
\end{figure}

\begin{figure}[t]
\begin{center}
\begin{tabular}{c}
\epsfig{figure=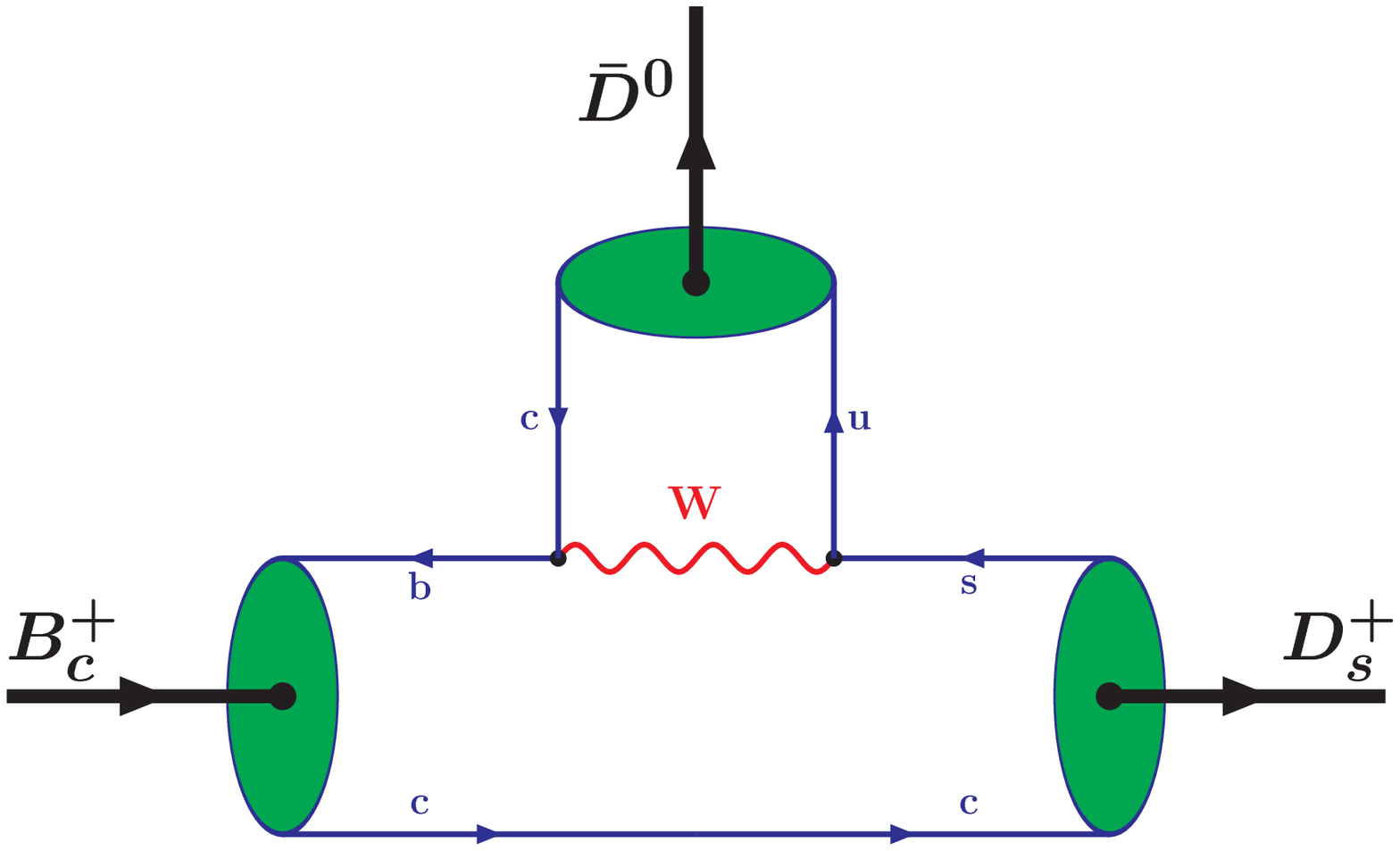,height=9cm}\\
\epsfig{figure=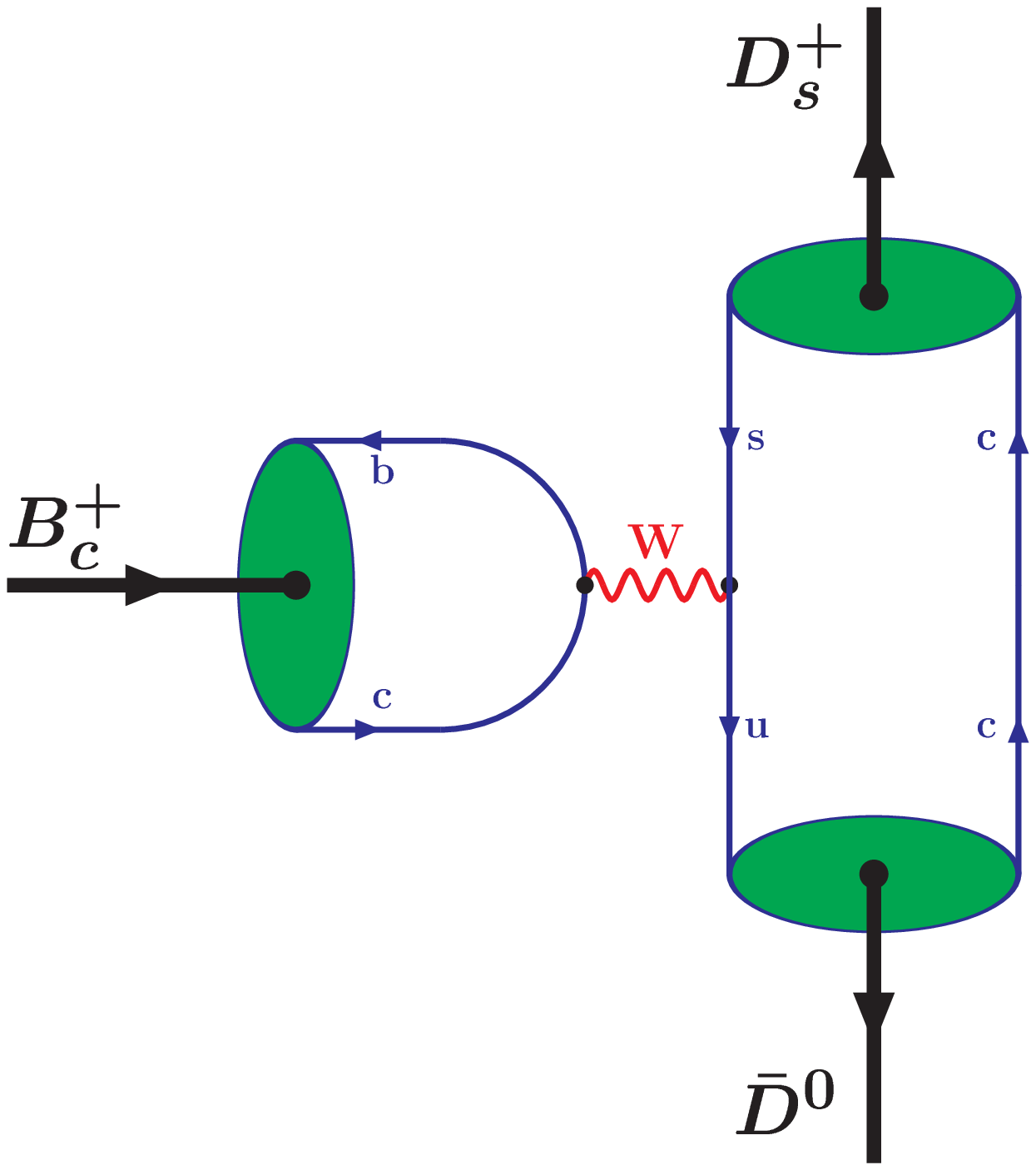,height=9cm}
\end{tabular}
\end{center}
\caption{Diagrams describing the decay $B_c\to D_s \overline {D^0}$.}
\label{fig:bc_dsdbar}
\end{figure}

\begin{figure}[t]
\begin{center}
\hspace*{-1cm}
\epsfig{figure=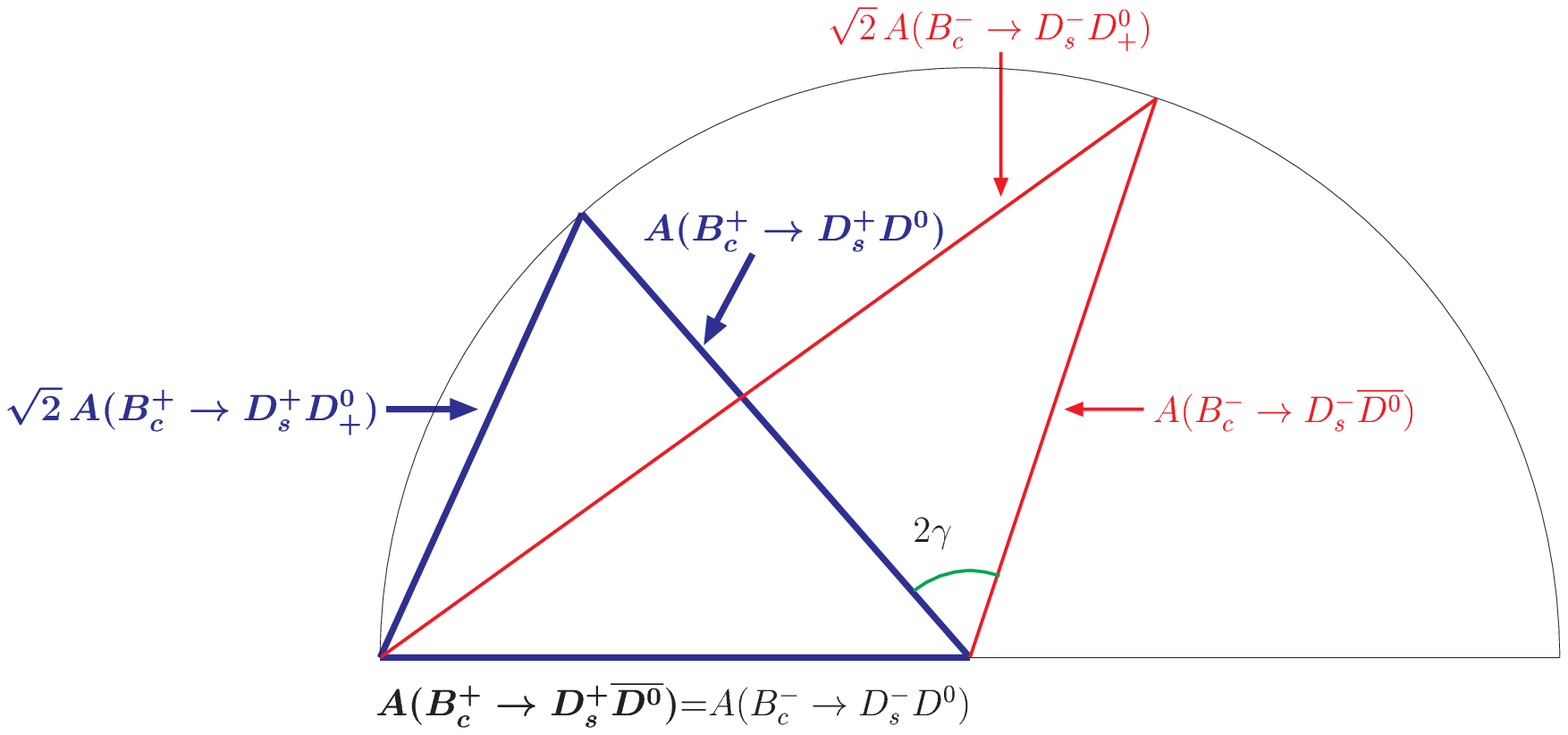,height=8cm} 
\end{center}
\caption{The amplitude triangles for the decays 
$B_c^\pm\to D^\pm_s\{D^0,\overline{D^0},D^0_+\}$.}
\label{fig:trn}
\end{figure}

In the case of the $B_c\to D_s D^0 (\overline {D^0})$ decays
the relevant amplitude relations can be  written 
in the form \cite{Fleischer:2000pp}

\begin{eqnarray}
\sqrt{2}\,A(B_c^+\to D_s^{+}D^0_+) &=&
A(B^+_c\to D_s^+ D^0)+ A(B^+_c\to D_s^+ \overline{D^0})\,,
\nonumber\\
&&\label{triangle}\\
\sqrt{2}\,A(B_c^-\to D_s^- D^0_+)  &=&
A(B^-_c\to D_s^- \overline{D^0})+A(B^-_c\to D_s^- D^0)\,,
\nonumber
\end{eqnarray}
where $|D^0_+\rangle=
\left(|D^0\rangle+|\overline{D^0}\rangle\right)/\sqrt{2}$
is a $CP$-even eigenstate.
The diagrams describing the decays $B^+_c\to D_s^+ D^0$
and $B^+_c\to D_s^+ \overline{D^0}$ are shown in Figs.~\ref{fig:bc_dsd}
and \ref{fig:bc_dsdbar}, respectively. The color-enhanced 
amplitude of  $B^+_c\to D_s^+ D^0$ can be seen to be proportional to
$V^{\dagger}_{ub}\cdot V_{cs}\approx 0.0029\cdot\exp(i\gamma)$.
At the same time the decay amplitude for $B^+_c\to D_s^+ \overline{D^0}$ 
is proportional to $V_{bc}\cdot V_{us}\approx 0.0088$ but
color-suppressed. Simple estimates made in \cite{Fleischer:2000pp} give

\begin{equation}
\label{bc-length}
\left|\frac{A(B^+_c\to D_s^+ D^0)}{A(B^+_c\to D_s^+ \overline{D^0})}\right|=
\left|\frac{A(B^-_c\to D_s^- \overline{D^0}}{A(B^-_c\to D_s^- D^0)}\right|=
{\cal O}(1).
\end{equation}
This implies that all sides of the amplitude triangles suggested in 
\cite{Fleischer:2000pp,Gronau:1991dp} have similar lengths as shown 
in  Fig.~\ref{fig:trn}.
It allows one to extract the magnitude of the weak CKM-phase $\gamma$
from the measurement of 
the $B_c^{\pm}\to D_s^{\pm}+(D^0,\overline{D^0},D^0_+)$ decay widths.
The method \cite{Gronau:1991dp} of the extraction of $\gamma$ 
from Eq.~(\ref{triangle})
is based on the parametrization of the amplitudes as

\begin{eqnarray}
A(B^+_c\to D_s^+ \overline{D^0})&=&A(B^-_c\to D_s^- D^0)
=|\bar A|\,e^{i\bar\delta},
\nonumber\\
&&\\
A(B^+_c\to D_s^+ D^0)&=& |A|\,e^{i\gamma}\,e^{i\delta},
\hspace{1cm}
A(B^-_c\to D_s^- \overline{D^0}))= |A|\,e^{-i\gamma}\,e^{i\delta}\,,
\nonumber
\end{eqnarray}
where $\delta$ and $\bar\delta$ are the strong final state interaction phases.
Introducing the notation $|A(B_c^+\to D_s^{+}D^0_+)|\equiv |A_+|$
and $|A(B_c^-\to D_s^- D^0_+)|\equiv |A_-|$   Eq.~(\ref{triangle})
can be rewritten as

\begin{eqnarray} 
|A_+|^2+|A_-|^2 &=&
|A|^2+|\bar A|^2+2\,|A|\,|\bar A|\,\cos\gamma\,\cos(\bar\delta-\delta)\,,
\nonumber\\
&&\\
|A_+|^2-|A_-|^2 &=& 2\,|A|\,|\bar A|\,\sin\gamma\,\sin(\bar\delta-\delta)\,.
\nonumber
\end{eqnarray}
The four solutions for $\sin\gamma$ are given by \cite{Gronau:1991dp}

\begin{equation}
\sin\gamma=\frac{1}{4\,|A|\,|\bar A|}
\left\{
\pm \sqrt{Y_{++}Y_{--}}\pm \sqrt{Y_{+-}Y_{-+}}
\right\}\,,
\label{solution}
\end{equation}
where $Y_{\pm+}=\left[|A|+|\bar A|\right]^2-2\,|A_\pm|^2$
and   $Y_{\pm-}=2\,|A_\pm|^2-\left[|A|-|\bar A|\right]^2$.
Thus, the measurements of the rates of the six decays
in Eq.~(\ref{triangle}) will determine the magnitude
of $\gamma$ with the four-fold ambiguity in Eq.~(\ref{solution}). 
The way to resolve the ambiguity was discussed in \cite{Gronau:1991dp}.

In contrast to $B_c\to D_sD$, the corresponding ratios for 
$B\to KD$ and  $B_c\to DD$ are \cite{Fleischer:2000pp}
\begin{eqnarray}
\label{BKD}
&&
\left|\frac{A(B^+\to K^+ D^0)}{A(B^+\to K^+ \overline{D^0})}\right|=
\left|\frac{A(B^-\to K^- \overline{D^0}}{A(B^-\to K^- D^0)}\right|=
{\cal O}(0.1),
\\
&&\nonumber\\
&&
\label{BcDD}
\left|\frac{A(B^+_c\to D^+ D^0)}{A(B^+_c\to D^+ \overline{D^0})}\right|=
\left|\frac{A(B^-_c\to D^- \overline{D^0}}{A(B^-_c\to D^- D^0)}\right|=
{\cal O}(0.1)
\end{eqnarray}
which can be seen to lead to squashed triangles which
are not very suited to measure $\gamma$. 

Some  estimates of the branching ratios have been obtained before
in \cite{Chang:1992pt}-\cite{AbdEl-Hady:1999xh} 
with widely divergent results.  We employ here 
a relativistic quark model \cite{Ivanov:1996pz} to provide
an independent evaluation of  these branching ratios. 

This model is based on an effective interaction Lagrangian 
which describes the coupling between hadrons and their constituent quarks.
For example, the coupling of the meson $H$ to its constituent quarks  $q_1$ 
and $\bar q_2 $ is given by the Lagrangian
\begin{equation}
\label{lag}
{\cal L}_{{\rm int}} (x)=g_H H(x) \int\!\! dx_1 \!\!\int\!\! dx_2
F_H (x,x_1,x_2) \bar q(x_1) \Gamma_H \lambda_H q(x_2)\,.
\end{equation}
Here, $\lambda_H$ and $\Gamma_H$ are  Gell-Mann and Dirac
matrices  which entail the flavor and spin quantum numbers
of the meson field $H(x)$. The shape of the vertex function $F_H$ 
can in principle be found from the  Bethe-Salpeter equation
as was done e.g. in \cite{Ivanov:1998ms}. However, we choose
a phenomenological approach where the vertex functions
are modelled by a simple form. 
The function $F_H$ must be invariant under 
the translation  $F_H(x+a,x_1+a,x_2+a)=F_H(x,x_1,x_2)$ and
should decrease quite rapidly in the Euclidean momentum space. 

In our previous papers \cite{Ivanov:2000aj} we omit 
a possible dependence of the vertex functions on external momenta
under calculation of the Feynman diagrams. This implies a dependence
on how loop momenta are routed through the diagram at hand.
In our last paper \cite{Faessler:2002ut} and in the present calculation
we employ  a particular form of the vertex function given by

\begin{equation}
\label{vertex}
F_H(x,x_1,x_2)=
\delta\biggl(x-\frac{m_1x_1+m_2x_2}{m_1+m_2}\biggr) \Phi_H((x_1-x_2)^2).
\end{equation}
where $m_1$ and $m_2$ are the constituent quark masses. 
The vertex function $F_H(x,x_1,x_2)$ evidently satisfies
the above translational invariance condition. 
We are able to make calculations explicitly without
any assumptions concerning the choice of loop momenta.

The coupling constants $g_H$ in Eq.~(\ref{lag}) are determined  
by the so called {\it compositeness condition} proposed in \cite{SWH} 
and extensively used in \cite{Efimov:zg}. The compositeness condition means 
that the renormalization constant of the meson field is set equal to zero

\begin{equation}
\label{z=0}
Z_H=1-\frac{3g^2_H}{4\pi^2}\tilde\Pi^\prime_H(m^2_H)=0
\end{equation}
where $\tilde\Pi^\prime_H$ is the derivative of the meson mass operator.
In physical terms the compositeness condition means that the meson
is composed of a quark and antiquark system.
For the pseudoscalar and vector mesons treated in this paper
one has

\begin{eqnarray*}
\tilde\Pi'_P(p^2)&=& \frac{1}{2p^2}\,p^\alpha\frac{d}{p^\alpha}\,
\int\!\! \frac{d^4k}{4\pi^2i} \tilde\Phi^2_P(-k^2)
\\ 
&&{\rm tr} \biggl[\gamma^5 S_1(\not\! k+w_{21} \not\!p) \gamma^5 
                         S_2(\not\! k-w_{12} \not\!p) \biggr] 
\\
&&\\
\tilde\Pi'_V(p^2)&=&
\frac{1}{3}\biggl[g^{\mu\nu}-\frac{p^\mu p^\nu}{p^2}\biggr] 
\frac{1}{2p^2}\,p^\alpha\frac{d}{p^\alpha}\,
\int\!\! \frac{d^4k}{4\pi^2i} \tilde\Phi^2_V(-k^2)
\\
&&
{\rm tr} \biggl[\gamma^\nu S_1(\not\! k+w_{21} \not\!p) \gamma^\mu 
                           S_2(\not\! k-w_{12}\not\! p)\biggr]
\end{eqnarray*}
where $w_{ij}=m_j/(m_i+m_j)$.

The leptonic decay constant $f_P$ is calculated from

\begin{eqnarray*}
&&
\frac{3g_P}{4\pi^2} \,\int\!\! \frac{d^4k}{4\pi^2i} 
\tilde\Phi_P(-k^2) 
{\rm tr} \biggl[O^\mu S_1(\not\! k+w_{21} \not\!p) \gamma^5 
                        S_2(\not\! k-w_{12} \not\!p) \biggr]
=f_P\,p^\mu.
\\
&&\\
&&
\frac{3g_V}{4\pi^2} \,\int\!\! \frac{d^4k}{4\pi^2i} 
\tilde\Phi_V(-k^2) 
{\rm tr} \biggl[O^\mu S_1(\not\! k+w_{21} \not\!p) \gamma
\cdot\epsilon_V S_2(\not\! k-w_{12} \not\!p) \biggr]
=\frac{1}{m_V}f_V\,\epsilon_V^\mu\,,
\end{eqnarray*}
The transition form factors  $P(p_1)\to P(p_2)(V(p_2))$ 
are calculated from the Feynman integrals

\begin{eqnarray}
&&
\frac{3g_P g_{P'}}{4\pi^2} \,\int\!\! \frac{d^4k}{4\pi^2i}
\tilde\Phi_P(-(k+w_{13}\,p_1)^2)\,
\tilde\Phi_{P'}(-(k+w_{23}\,p_2)^2)
\label{PP}\\
&&
\cdot{\rm tr} \biggl[S_2(\not\! k+\not\! p_2) O^\mu 
 S_1(\not\! k+\not\! p_1)\gamma^5 S_3(\not\! k)\gamma^5\biggr]
=F_+(q^2) P^\mu+F_-(q^2) q^\mu\,,
\nonumber\\
&&\nonumber\\
&&
\frac{3g_P g_{V}}{4\pi^2} \,\int\!\! \frac{d^4k}{4\pi^2i}
\tilde\Phi_P(-(k+w_{13}\,p_1)^2)\,
\tilde\Phi_{V}(-(k+w_{23}\,p_2)^2)
\label{PV}\\
&&
\cdot{\rm tr} \biggl[S_2(\not\! k+\not\! p_2) O^\mu 
 S_1(\not\! k+\not\! p_1)\gamma^5 S_3(\not\! k)\gamma\cdot\epsilon_V\biggr]
=\frac{(\epsilon_{V})_\nu}{m_P+m_{V}}\,
\nonumber\\
&&\nonumber\\
&&
\left\{-g^{\mu\nu}\,Pq\,A_0(q^2)+P^\mu P^\nu\,A_+(q^2)
+q^\mu P^\nu\,A_-(q^2)
+i\varepsilon^{\mu\nu\alpha\beta} P_\alpha q_\beta \,V(q^2)\right\}\,,
\nonumber
\end{eqnarray}
where 
$O^\mu=\gamma^\mu(1-\gamma^5)$.
We use the local quark propagators
$
S_i(\not\! k)=1/(m_i-\not\!  k)
$
where $m_i$ is  the constituent quark mass. As discussed in
\cite{Ivanov:2000aj,Faessler:2002ut}, we assume that 
$ 
m_H<m_{1}+m_{2}
$
in order to avoid the appearance of imaginary parts in  the physical
amplitudes. This holds true for the light pseudoscalar mesons 
but is no longer true for the light vector mesons. 
We shall therefore employ identical  masses for the
pseudoscalar mesons and the vector mesons
in our matrix element calculations but use physical masses 
in the phase space calculation. This is quite a reliable approximation for the
heavy mesons, e.g. $D^\ast$ and $B^\ast$ whose masses are almost
the same as the  $D$ and $B$, respectively. However,  for the light mesons
this approximation is not so good 
since the  $K^\ast(892)$ has a mass much larger than the  $K(494)$.   
For this reason we exclude the light vector mesons
from our considerations.
The fit values for the constituent quark masses are taken from our
papers \cite{Ivanov:2000aj,Faessler:2002ut}  and are given in 
Eq.~(\ref{fitmas}). 

\begin{equation}
\begin{array}{ccccc}
     m_u        &      m_s        &      m_c       &     m_b        &   \\
\hline
$\ \ 0.235\ \ $ & $\ \ 0.333\ \ $ & $\ \ 1.67\ \ $ & $\ \ 5.06\ \ $ & 
$\  {\rm GeV} $\\
\end{array}
\label{fitmas}
\end{equation}

We employ a Gaussian for the vertex functuon 
$\tilde\Phi_H(k_E^2)=\exp(-k_E^2/\Lambda^2_H)$
and determine the size parameters $\Lambda^2_H$ by a fit to
the experimental data, when available, or to lattice simulations 
for the leptonic decay constants.
The numerical values for $\Lambda_H$ are 

\begin{equation}\label{lambda}
\begin{array}{cccccccc}
\Lambda_\pi & \Lambda_K & \Lambda_D & \Lambda_{D_s} &
 \Lambda_B & \Lambda_{B_s} & \Lambda_{B_c} & \\ 
\hline 
$\ \ 1.00\ \ $ & $\ \ 1.60\ \ $ & $\
\ 1.70\ \ $ & $\ \ 1.70\ \ $ & $\ \ 2.00\ \ $ &
$\ \ 2.00\ \ $ &$\ \ 2.05\ \ $ & $\ {\rm GeV} $  \\
\end{array}
\end{equation}

We have used the technique outlined in our previous papers
 \cite{Ivanov:2000aj,Faessler:2002ut} 
for the numerical  evaluation  of the Feynman integrals 
in Eqs.~(\ref{PP}) and (\ref{PV}).
The results of our numerical calculations are well represented
by the parametrization
\begin{equation}\label{approx}
F(s)=\frac{F(0)}{1-a \hat s+b \hat s^2}\,
\end{equation} 
with $\hat s=q^2/m^2_{B_c}$.
Using such a  parametrization facilitates further integrations.
The values of $F(0)$, $a$ and $b$ are listed  in Table~\ref{tab:ff}.
The calculated values of the leptonic decay constants are given
in Eq.~(\ref{leptonic}). They agree with the available
experimental data and the results of the lattice simulations. 

\begin{equation}
\def\arraystretch{1.5}
\begin{array}{ccccccccc}
  f_{K^+} & f_{D^0} & f_{D^{\ast\, 0}}  & 
  f_{D_s} & f_{D_s^{\ast}}    & f_B     &  f_{B_c} &           \\   
\hline
$\ \ 0.161\ \ $  & $\ \ 0.215\ \ $ & 
$\ \ 0.348\ \ $  & $\ \ 0.222\ \ $ & $\ \ 0.329\ \ $ &
$\ \ 0.180\ \ $  & $\ \ 0.398\ \ $ & $\ \ {\rm GeV} $  \\
\end{array}
\label{leptonic}
\end{equation}

The relevant effective Hamiltonian for the decays 
$B_c\to D_s\overline {D^0}$ and  $B_c\to D_s D^0$ 
is written as 

\begin{eqnarray}
\label{eff}
H_{\rm eff}&=&-\frac{G_F}{\sqrt{2}}\,
\left\{
C_1(\mu)\left(V_{cs}V^\dagger_{ub}\cdot (\bar b u)_{V-A}(\bar c s)_{V-A}
             +V_{us}V^\dagger_{cb}\cdot (\bar b c)_{V-A}(\bar u s)_{V-A}\right)
\right.
\nonumber\\
&+&
\left.
C_2(\mu)\left(V_{cs}V^\dagger_{ub}\cdot (\bar b s)_{V-A}(\bar c u)_{V-A}
             +V_{us}V^\dagger_{cb}\cdot (\bar b s)_{V-A}(\bar u c)_{V-A}\right)
\right\}
\end{eqnarray}
where $V-A$ refers to $O^\mu=\gamma^\mu(1-\gamma^5)$.
We use the numerical values of the Wilson coefficients at
the renormalization scale $\mu=m_{b,\rm pole}$ given by
$C_1=1.107$ and $C_2=-0.248$ as given in \cite{Ali:2002jg}. 
Note that we interchange the labeling
$1\leftrightarrow 2$ of the coefficients to be consistent
with the papers \cite{Chang:1992pt}-\cite{AbdEl-Hady:1999xh}.

Straightforward calculation of the matrix elements of the decays 
$B_c\to D_s\overline {D^0} (D_s D^0)$ by using the effective Hamiltonian 
in Eq.(\ref{eff}) reproduces the result of the factorization method.
We have
 
\begin{eqnarray}
A(B^+_c\to D^+_s D^0) &=& \frac{G_F}{\sqrt{2}}\cdot V_{ub}^\dagger V_{cs} 
\label{dsd}\\
&\times&
\left\{
a_1\,\left[f_+^{B_c D}(m^2_{D_s})\,(m^2_{B_c}-m^2_{D^0})
          +f_-^{B_c D}(m^2_{D_s})\, m^2_{D_s}\right]\cdot f_{D_s}
\right.
\nonumber\\
&&\nonumber\\
&+&
\left.
a_2\,\left[f_+^{B_c D_s}(m^2_{D^0})\,(m^2_{B_c}-m^2_{D_s})   
          +f_-^{B_c D_s}(m^2_{D^0})\, m^2_{D^0}\right]\cdot f_{D^0}
\right\}\,,
\nonumber\\
&&\nonumber\\
&&\nonumber\\
A(B^+_c\to D^+_s \overline {D^0}) &=& 
\frac{G_F}{\sqrt{2}}\cdot V_{bc}^\dagger V_{us}
\label{dsdbar}\\
&\times&
a_2\,\left[f_+^{B_c D_s}(m^2_{D^0})\,(m^2_{B_c}-m^2_{D_s})   
          +f_-^{B_c D_s}(m^2_{D^0})\, m^2_{D^0}\right]\cdot f_{D^0}
\nonumber\\
&+&
{\rm annihilation \,\,\, channel}\nonumber
\end{eqnarray}
where $a_1=C_1+\xi C_2$ and  $a_2=C_2+\xi C_1$ with $\xi=1/N_c$.
As usual we put the QCD color factor $\xi=0$ according to
$1/N_c$-expansion. Also we drop the annihilation processes
from the consideration. Note that the calculation
of the matrix elements of the nonleptonic decays involving
the vector D-mesons in the final states proceed in a similar way.
We extend our analysis to the semileptonic and nonleptonic 
decays of $B$-meson.

For numerical evaluation we have used the set of the parameters:
$m_{B^+}=5.279$ GeV, $\tau_{B^+}=1.655$ ps,
$m_{B_c}=6.4$ GeV, $\tau_{B_c}=0.46$ ps, 
$a_1|_{\xi=0}=1.107$, $a_2|_{\xi=0}=-0.248$ and

\begin{equation}
\def\arraystretch{1.5}
\begin{array}{cccccc}
 |V_{ud}| &  |V_{us}| & |V_{ub}|  & |V_{cd}| &  |V_{cs}| & |V_{bc}|   \\   
\hline
$\ \ 0.98\ \ $ & $\ \ 0.22\ \ $ & $\ \ 0.003\ \ $ & $\ \ 0.22\ \ $ & 
$\ \ 0.98\ \ $ & $\ \ 0.040\ \ $ \\ 
\end{array}
\label{ckm}
\end{equation}

First, to illustrate the quality of our calculations,
we list some branching ratios of the $B$-meson decays
in Table~\ref{tab:fit} and compare them with 
the experimental data. 
The exclusive nonleptonic decay widths of the $B$ and $B_c$ mesons
for arbitrary values of $a_1$ and $a_2$ are listed in Table~\ref{tab:br1}
whereas their branching ratios for $a_1=1.107$ and $a_2=-0.248$
are given in Table~\ref{tab:br2}. One can see that as it was expected
the magnitudes of the branching ratios of the decays $B_c\to D_s D^0$
and $B_c\to D_s \overline{ D^0}$ are very close to each other. It gives us
hope that they can be measured in the forthcoming experiments.
Finally, in Table~\ref{tab:br3} we compare  our results with 
the results of other studies. One can see that there are 
quite large differences between the predictions of
the different models. 

This work was completed while M.A.I. and O.N.P. visited the University 
of Mainz. M.A.I. appreciates the partial support by
Graduiertenkolleg ``Eichtheorien'',
the Russian Fund of Basic Research 
grant No. 01-02-17200 and the Heisenberg-Landau Program.
O.N.P. thanks DAAD and Federal Russian Program ''Integracia''
Grant No. 3057.

\begin{table}[b]
\caption{
Form factors for $B_{c}^{+}\rightarrow D^0(D^{\ast 0})$
and $B_{c}^{+}\rightarrow D_s^+(D_s^{\ast +})$ transitions.
Form factors are {\it approximated} by the form
$F(q^2)=F(0)/(1-a\,\hat s+b\,\hat s^2)$ with $\hat s=q^2/m_{B_c}^2$.
}
\vspace{0.3cm}
\begin{center}
\def\arraystretch{1.5}
\begin{tabular}{|c|c|c|c||c|c|c|}
\hline
   & \multicolumn{3}{c||} { $B_{c}^{+}\rightarrow D^0(D^{\ast 0})     $} 
   & \multicolumn{3}{c|} { $B_{c}^{+}\rightarrow D_s^+(D_s^{\ast +}) $} \\
\hline
        & $F(0)$& $a$  & $b$  &$F(0)$ & $a$  & $b$   \\ 
\hline\hline
$F_{+}$ & 0.189 & 2.47 & 1.62 & 0.194 & 2.47 & 1.61  \\ \hline
$F_{-}$ &-0.194 & 2.43 & 1.54 &-0.183 & 2.43 & 1.53  \\ \hline\hline
$A_{0}$ & 0.284 & 1.30 & 0.15 & 0.312 & 1.40 & 0.16  \\ \hline
$A_{+}$ & 0.158 & 2.15 & 1.15 & 0.168 & 2.21 & 1.19  \\ \hline
$A_{-}$ &-0.328 & 2.40 & 1.51 &-0.329 & 2.41 & 1.51  \\ \hline
$V    $ & 0.296 & 2.40 & 1.49 & 0.298 & 2.41 & 1.49 \\ \hline
\end{tabular}
\label{tab:ff}
\end{center}
\end{table}

\begin{table}[ht]
\caption{
Comparison of some branching ratios 
of the $B$-meson decays
with the available experimental data. 
}
\vspace{0.3cm}
\begin{center}
\def\arraystretch{1.5}
\begin{tabular}{|c|c|c|}
\hline
                              & This work & PDG \cite{Hagiwara:fs}   \\
\hline\hline
$B^+\to \overline {D^0}\,e^+\nu$ & 0.024     & 0.0215$\pm$0.0022 \\
\hline
$B^+\to \overline {D^{\ast\,0}}\,e^+\nu$ & 0.056     & 0.053$\pm$0.008 \\
\hline\hline
$  B^+\to K^+\, \overline {D^0}$ & $2.8\cdot 10^{-4}$ & 
                            $(2.9\pm 0.8)\cdot 10^{-4}$ \\
\hline\hline
$ B^+\to D_s^+\, \overline {D^0}$   & $0.013$ & $0.013\pm 0.004$ \\  
\hline
$ B^+\to D_s^+\,\overline{D^{\ast 0}}$    & $0.008$ & $0.012\pm 0.005$ \\ 
\hline
$ B^+\to D_s^{\ast +}\, \overline{D^0}$   & $0.019$ & $0.009\pm 0.004$ \\  
\hline
$ B^+\to D_s^{\ast +}\,\overline{D^{\ast 0}}$ & $0.046$ &  
                                           $0.027\pm 0.010$ \\  
\hline 
\end{tabular}
\label{tab:fit}
\end{center}
\end{table}

\begin{table}[ht]
\caption{ Exclusive nonleptonic  decay widths 
of the $B$ and$B_c$ mesons in $10^{-15}$ GeV.}
\begin{center}
\vspace{0.3cm}
\def\arraystretch{1.5}
\begin{tabular}{|c|c||c|c|}
\hline
$ B^+\to K^+\, \overline {D^0}$               & $(0.364\,a_1+0.286\,a_2)^2$ &  
$ B^+\to K^+\, D^0$          & $ 0.00915\,a_2^2 $ \\
\hline
$ B^+\to K^+\,\overline{  D^{\ast 0}}$        & $(0.342\,a_1+0.442\,a_2)^2$ &  
$ B^+\to K^+\, D^{\ast 0}$   & $ 0.0219\,a_2^2 $ \\ 
\hline\hline
$ B^+\to D_s^+\, \overline{ D^0}$               & $4.367\,a_1^2$ &  
 & \\
\hline
$ B^+\to D_s^+\,\overline{  D^{\ast 0}}$        & $2.707\,a_1^2$ &  
  &  \\ 
\hline
$ B^+\to D_s^{\ast +}\, \overline{  D^0}$       & $6.300\,a_1^2$ &  
  &   \\
\hline
$ B^+\to D_s^{\ast +}\,\overline { D^{\ast 0}}$ & $14.84\,a_1^2$ &  
 & \\ 
\hline\hline
$ B_c^+\to D^+\, D^0$               & $(0.0147\,a_1+0.0146\,a_2)^2$ &  
$ B_c^+\to D^+\, \overline{ D^0}$          & $ 0.753\,a_2^2 $ \\
\hline
$ B_c^+\to D^+\, D^{\ast 0}$        & $(0.0107\,a_1+0.0234\,a_2)^2$ &  
$ B_c^+\to D^+\, \overline{ D^{\ast 0}}$   & $ 1.925\,a_2^2 $ \\ 
\hline
$ B_c^+\to D^{\ast +}\,  D^0$       & $(0.0233\,a_1+0.0106\,a_2)^2$ &  
$ B_c^+\to D^{\ast +}\, \overline{ D^0}$   & $ 0.399\,a_2^2 $  \\
\hline
$ B_c^+\to D^{\ast +}\, D^{\ast 0}$ & $(0.0235\,a_1+0.0235\,a_2)^2$ &  
$ B_c^+\to D^{\ast +}\, \overline{ D^{\ast 0}}$ & $ 1.95\,a_2^2 $ \\ 
\hline\hline
$ B_c^+\to D_s^+\, D^0$               & $(0.0689\,a_1+0.0672\,a_2)^2$ &  
$ B_c^+\to D_s^+\, \overline{ D^0}$          & $ 0.0405\,a_2^2 $ \\
\hline
$ B_c^+\to D_s^+\, D^{\ast 0}$        & $(0.0503\,a_1+0.106\,a_2)^2$ &  
$ B_c^+\to D_s^+\, \overline{ D^{\ast 0}}$   & $ 0.101\,a_2^2 $ \\ 
\hline
$ B_c^+\to D_s^{\ast +}\,  D^0$       & $(0.101\,a_1+0.0498\,a_2)^2$ &  
$ B_c^+\to D_s^{\ast +}\, \overline{ D^0}$   & $ 0.0222\,a_2^2 $  \\
\hline
$ B_c^+\to D_s^{\ast +}\, D^{\ast 0}$ & $(0.104\,a_1+0.110\,a_2)^2$ &  
$ B_c^+\to D_s^{\ast +}\, \overline{ D^{\ast 0}}$ & $ 0.109\,a_2^2 $ \\ 
\hline
\end{tabular}
\label{tab:br1}
\end{center}
\end{table}

\begin{table}[ht]
\caption{ Branching ratios of some  nonleptonic  decay widths 
of the $B$ and$B_c$ mesons in calculated for $a_1=1.107$ and 
$a_2=-0.248$.}
\vspace{0.3cm}
\begin{center}
\def\arraystretch{1.5}
\begin{tabular}{|c|c||c|c|}
\hline
$ B^+\to K^+\, \overline{ D^0}$               & $2.76\cdot 10^{-4}$ &  
$ B^+\to K^+\, D^0$          & $1.41\cdot 10^{-6} $ \\
\hline
$ B^+\to K^+\,\overline{  D^{\ast 0}}$        & $1.82\cdot 10^{-4}$ &  
$ B^+\to K^+\, D^{\ast 0}$   & $3.38\cdot 10^{-6}  $ \\ 
\hline\hline
$ B_c^+\to D^+\, D^0$               & $1.11\cdot 10^{-7}$ &  
$ B_c^+\to D^+\, \overline{ D^0}$          & $3.24\cdot 10^{-5}$ \\
\hline
$ B_c^+\to D^+\, D^{\ast 0}$        & $0.25\cdot 10^{-7}$ &  
$ B_c^+\to D^+\, \overline{ D^{\ast 0}}$   & $8.28\cdot 10^{-5}$ \\ 
\hline
$ B_c^+\to D^{\ast +}\,  D^0$       & $3.76\cdot 10^{-7}$ &  
$ B_c^+\to D^{\ast +}\, \overline{ D^0}$   & $1.71\cdot 10^{-5}$  \\
\hline
$ B_c^+\to D^{\ast +}\, D^{\ast 0}$ & $2.84\cdot 10^{-7}$ &  
$ B_c^+\to D^{\ast +}\, \overline{ D^{\ast 0}}$ & $ 8.38\cdot 10^{-5} $ \\ 
\hline\hline
$ B_c^+\to D_s^+\, D^0$               & $2.48\cdot 10^{-6}$ &  
$ B_c^+\to D_s^+\, \overline{ D^0}$          & $1.74\cdot 10^{-6}$ \\
\hline
$ B_c^+\to D_s^+\, D^{\ast 0}$        & $0.60\cdot 10^{-6}$ &  
$ B_c^+\to D_s^+\, \overline{ D^{\ast 0}}$   & $4.34\cdot 10^{-6}$ \\ 
\hline
$ B_c^+\to D_s^{\ast +}\,  D^0$       & $6.88\cdot 10^{-6}$ &  
$ B_c^+\to D_s^{\ast +}\, \overline{ D^0}$   & $0.95\cdot 10^{-6} $  \\
\hline
$ B_c^+\to D_s^{\ast +}\, D^{\ast 0}$ & $5.41\cdot 10^{-6}$ &  
$ B_c^+\to D_s^{\ast +}\, \overline{ D^{\ast 0}}$ & $4.69\cdot 10^{-6} $ \\ 
\hline
\end{tabular}
\label{tab:br2}
\end{center}
\end{table}

\begin{table}[ht]
\caption{ 
Exclusive nonleptonic  decay widths of the $B_c$ meson
in units of $10^{-15}$ GeV. \newline
Comparison with other studies.}
\vspace{0.3cm}
\begin{center}
\def\arraystretch{1.5}
\begin{tabular}{|c|c|c|c|c|c|c|}
\hline
Process  & This paper & \cite{Chang:1992pt} & \cite{Liu:1997hr} & 
\cite{Colangelo:1999zn} & \cite{AbdEl-Hady:1999xh} & \cite{Gouz:2002kk}\\
\hline\hline
$ B_c^+\to D_s^+\, \overline{ D^0}$        & $ 0.0405\,a_2^2 $ &
                                             $ 0.0340\,a_2^2 $ &
                                             $ 0.168 \,a_2^2 $ &
                                             $ 0.01  \,a_2^2 $ &
                                             $ 0.0415\,a_2^2 $ &
                                             $ 0.176 \,a_2^2 $ \\
\hline
$ B_c^+\to D_s^+\, \overline{ D^{\ast 0}}$ & $ 0.101 \,a_2^2 $ &
                                             $ 0.0354\,a_2^2 $ &
                                             $ 0.143 \,a_2^2 $ &
                                             $ 0.009 \,a_2^2 $ &
                                             $ 0.0495\,a_2^2 $ &
                                             $ 0.260 \,a_2^2 $ \\ 
\hline
$ B_c^+\to D_s^{\ast +}\, \overline{ D^0}$ & $ 0.0222\,a_2^2 $ &
                                             $ 0.0334\,a_2^2 $ &
                                             $ 0.0658\,a_2^2 $ &
                                             $ 0.087 \,a_2^2 $ &
                                             $ 0.0201\,a_2^2 $ & 
                                             $ 0.166 \,a_2^2 $ \\ 
\hline
$ B_c^+\to D_s^{\ast +}\, \overline{ D^{\ast 0}}$ 
                                           & $ 0.109 \,a_2^2 $ &
                                             $ 0.0564\,a_2^2 $ &
                                             $ 0.128 \,a_2^2 $ &
                                             $ 0.15  \,a_2^2 $ &
                                             $ 0.0597\,a_2^2 $ &
                                             $ 0.951 \,a_2^2 $ \\
\hline
\end{tabular}
\label{tab:br3}
\end{center}
\end{table}

\clearpage

\end{document}